# Experimental estimations of viscoelastic properties of multilayer damped plates in broad-band frequency range

KeremEge[a)]
Laboratoire Vibrations Acoustique, INSA de Lyon
25 bis avenue Jean Capelle, 69621 Villeurbanne Cedex, FRANCE

Thibault Boncompagne[b)]
GMPPA, Site de Plasturgie de l'INSA de Lyon, BP 807, 01108 OYONNAX CEDEX

Bernard Laulagnet[c)]
Laboratoire Vibrations Acoustique, INSA de Lyon
25 bis avenue Jean Capelle, 69621 Villeurbanne Cedex, FRANCE

Jean-Louis Guyader[d)]
Laboratoire Vibrations Acoustique, INSA de Lyon
25 bis avenue Jean Capelle, 69621 Villeurbanne Cedex, FRANCE

**Regarding lightweighting structures for aeronautics, automotive or construction applications, the level of performance of solutions proposed in terms of damping and isolation is fundamental. Hence multilayered plate appears as an interesting answer if damping performances are properly optimized. In this paper, a novel modal analysis method (Ege et al, JSV 325 (4-5), 2009) is used to identify viscoelastic properties (loss factors, Young's modulus) of "polyethylene thermoplastic / aluminum" bilayer plates. The thermoplastic is chosen for its high loss factors and relative low mass. The experimental method consists in a high-resolution technique (ESPRIT algorithm) which allows precise estimations of the viscoelastic properties even in frequency domains with high modal overlap (high damping or modal density). Experimental loss factors estimated from impact hammer excitations on the free-free plates highly corresponds with two theoretical estimations. In the first model (Guyader&Lesueur, JSV 58(1), 1978) the calculation is based on multilayered plates equations and use wave propagation analysis ; in the second one (Laulagnet&Guyader, JASA 96(1), 1994) the thickness deformation solving Navier's equations is allowed. Results on several plates with several thicknesses of thermoplastics are given and compared with the models, demonstrating the validity of the approach.**

[a)] email: kerem.ege@insa-lyon.fr
[b)] email: thibault.boncompagne@insa-lyon.fr
[c)] email: bernard.laulagnet@insa-lyon.fr
[d)] email: jean-louis.guyader@insa-lyon.fr

## 1 INTRODUCTION

The present communication is concerned with the study of damping properties of multilayer plates in the context of vibro-acoustics applications. In a first part, the experimental protocol is presented; using a novel modal analysis method (Ege et al, JSV 325 (4-5), 2009) the viscoelastic properties (Young's modulus, loss factors) of several bilayer plates of different thicknesses are identified up to several kHz even in frequency domains with high modal overlap (high damping or modal density). Results are presented in section 3, and compared with two analytical models in section 4.

## 2 EXPERIMENTAL PROTOCOL

### 2.1 Bilayer plates: choice of polymer and manufacture

Each plate is constituted of a first layer in aluminum (of thickness 2mm), on which a second layer of damping material is glued with different thicknesses. The selection of the polymer material constituting the second layer was done using Ashby diagrams. The purpose was to select a polymer material with a high loss factor $\eta$, a relatively low density $\rho$, and a high Young's modulus $E$. The two Ashby diagrams presenting Young's modulus versus density and loss factor versus Young's modulus are presented in Fig.1 and Fig. 2.

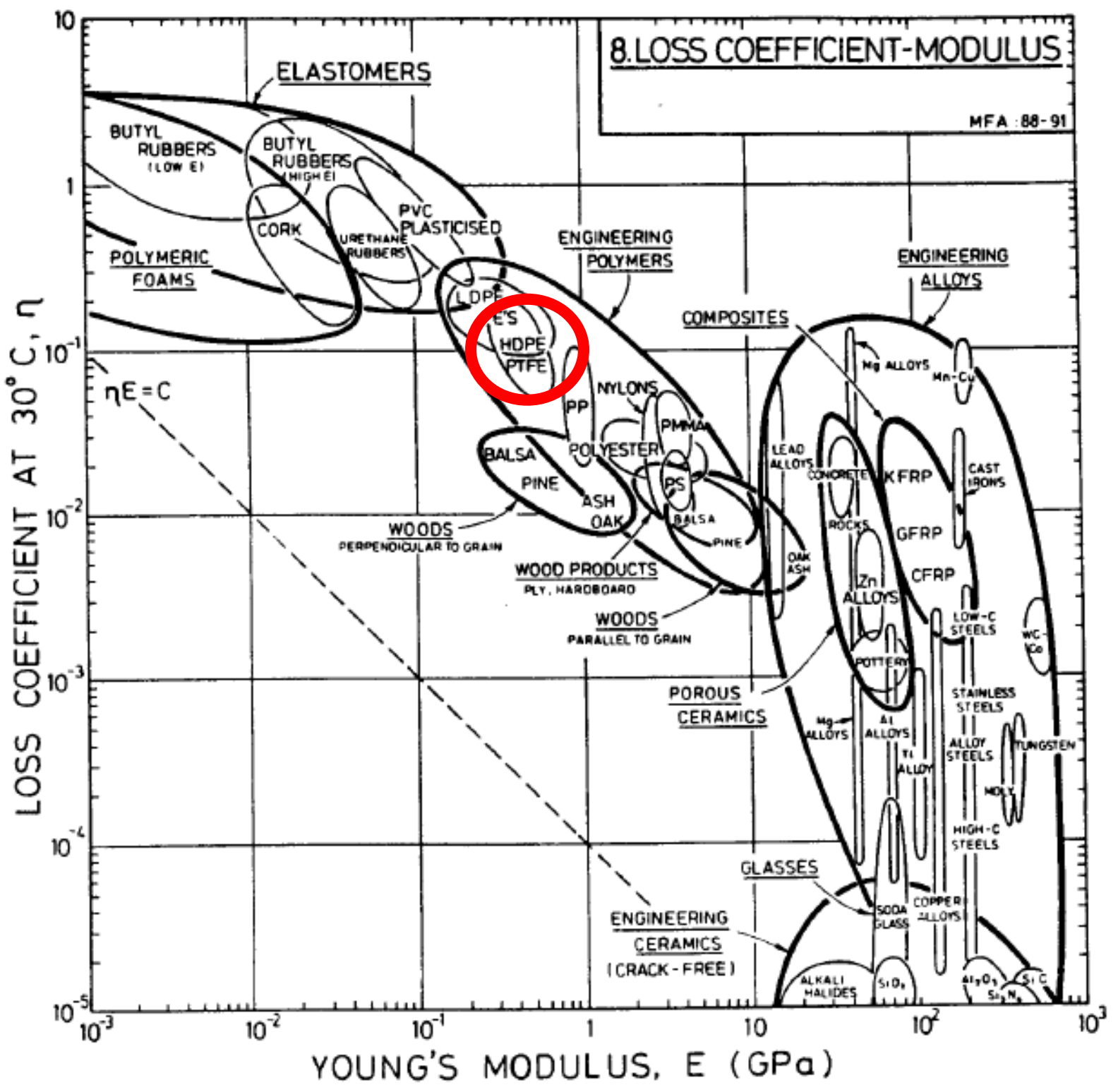

*Fig. 1–Loss factor $\eta$ versus Young's modulus$E$diagram, after Ashby[1]. HDPE is indicated in red circle.*

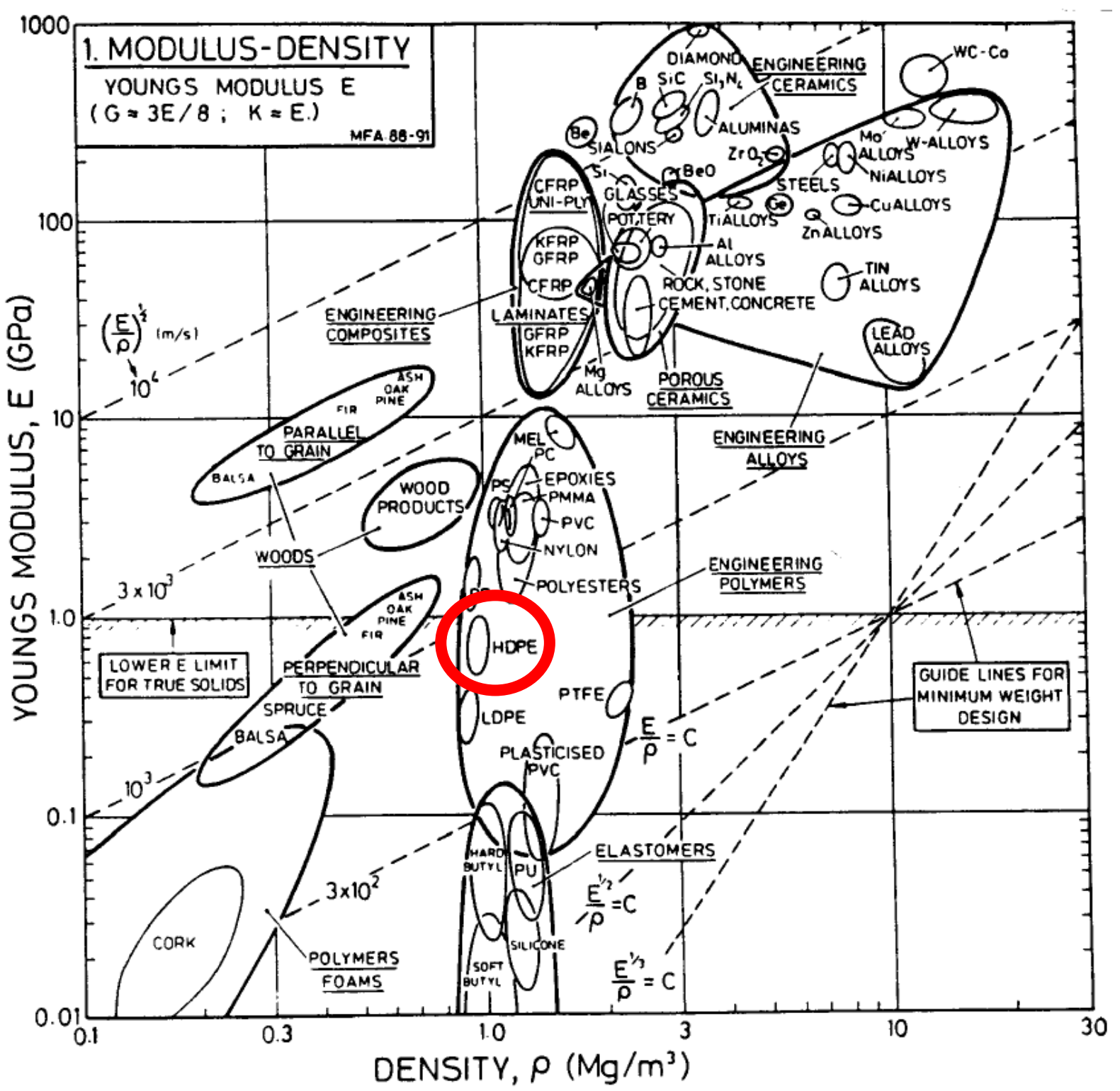

*Fig. 2– Young's modulus E versus density ρ diagram, after Ashby*[1].

The indicator $A = E\,\eta/\rho$ is defined and calculated for typical polymers. Results are given in Fig.3. It appears that the High-Density PolyEthylene (HDPE)maximizes this indicator (about three times more than for the other polymers). Therefore, this polymer is retained in our study for the damping layer. The mechanical characteristics of the HDPE are: $E$=850 MPa; $\rho$=945 kg.m$^{-3}$ and $\eta$=15%. These Young's modulus and loss factors were measured on a typical DMA tester (at low frequencies).

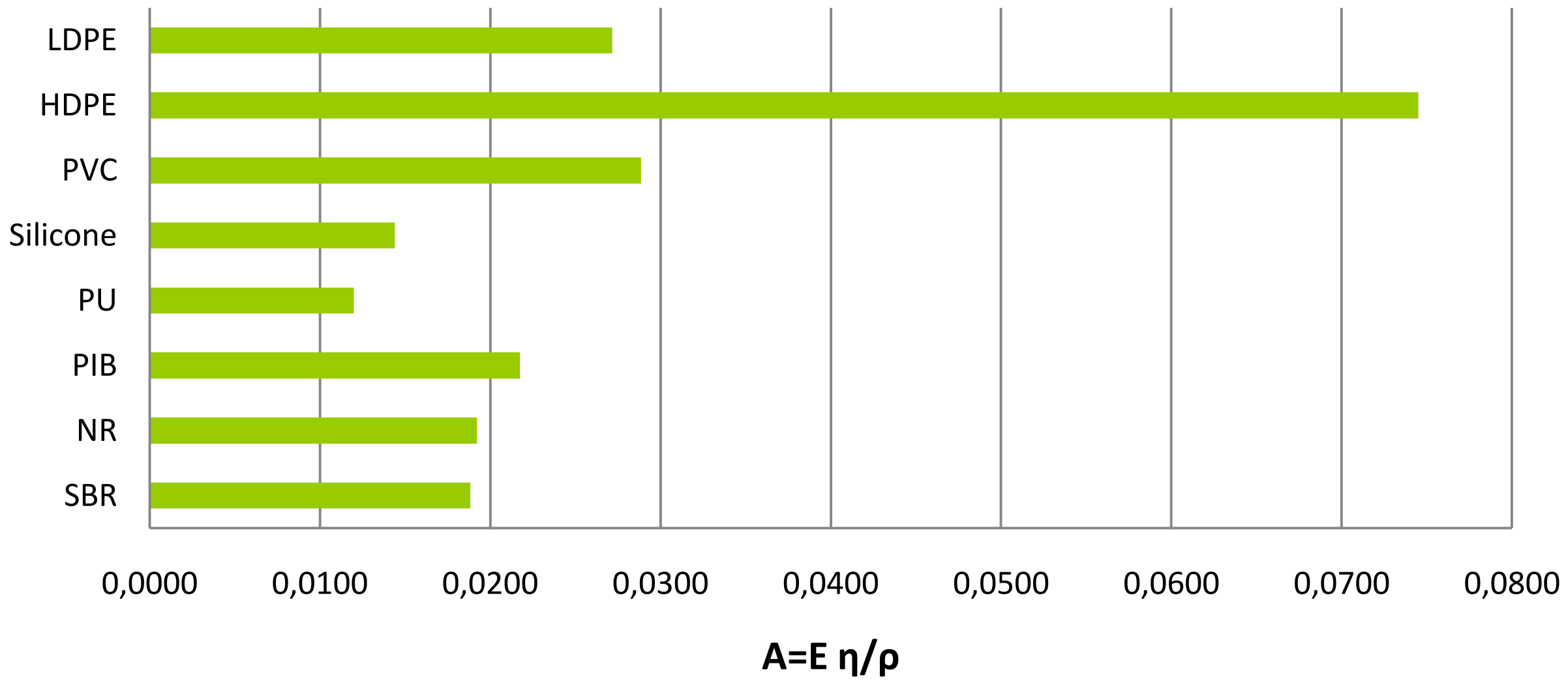

*Fig. 3 – Indicator $A = E\,\eta/\rho$for different polymers: Low-density polyethylene (LDPE); High-density polyethylene (HDPE); Polyvinyl chloride (PVC); Silicone; Polyurethane (PU); Polyisobutylene Rubber (PIB); Natural Rubber (NR); Styrene-butadiene rubber (SBR). HDPE maximizes this indicator (about three times more than for the other polymers).*

## 2.2 Vibratory measurements

Four bilayer plates were manufactured with different thicknesses of HDPE layer (0.5mm, 1mm, 1.5mm and 2mm) glued on a 2mm aluminum plate. The dimensions of the plates are $620\times420mm^2$. The experimental protocol consists in modal analyses of the free-free plates.
A pseudo-impulse force is applied by means of a small impact hammer. The acceleration is measured with an accelerometer. In all cases, boundary conditions are kept as close as possible to “free-free”, by suspending the plate from one of its corner. The point of excitation and the vibration measurements are made in the vicinity of another corner of the plate. Under the chosen boundary conditions, this location is not on any of the nodal lines.

## 2.3 ESPRIT method and signal processing

We used a novel modal analysis technique[2] based on ESPRIT algorithm[3] to estimate accurately the modal frequencies and loss factors of the bilayer plates. This high-resolution method assumes thatthe signal$s(t)$ is a sum of complex exponentials $x(t)$ (the modal signal to be determined) and white noise $\beta(t)$:

$$s(t) = x(t) + \beta(t) = \sum_{k=1}^{K} a_k e^{-\alpha_k t} e^{i(2\pi f_k t + \varphi_k)} + \beta(t) = \sum_{k=1}^{K} b_k z_k^t + \beta(t) \qquad (1)$$

where$K$ is the number of complex exponentials, $b_k = a_k e^{i\varphi_k}$ are the complex amplitudes (with $a_k$ and $\varphi_k$ the modal amplitudes and phases at the point of interest), and $z_k = e^{-\alpha_k t} e^{i\,2\pi f_k}$ the so-called poles (with $f_k$ the modal frequencies in Hz and $\alpha_k$ the modal damping factors in $s^{-1}$).
The modal damping factor $\alpha_k$ (also called modal decay constant in $s^{-1}$), the modal decay time $\tau_k$ (in s) and the modal loss factor $\eta_k$ (dimensionless) are related between them as follows:

$$\alpha_k = \frac{1}{\tau_k} = \frac{\eta_k \omega_k}{2} \quad , \quad \eta_k = \frac{\Delta f_{k,-3dB}}{f_k} = \frac{\alpha_k}{\pi f_k} \qquad (2)$$

where$\omega_k$ is the modal angular frequency (in $rad.s^{-1}$) and $\Delta f_{k,-3dB}$ the half-power modal bandwidth.

The rotational invariance property of the signalsubspace (see Roy *et al.*[3] for mathematical developments) is used to estimate the modal parameters: frequencies,damping factors and complex amplitudes. The dimensions of both subspacesmust be chosen *a priori* and the quality of the estimation depends significantlyon a proper choice for these parameters. The best choice for the dimension of themodal subspace is the number of *complex* exponentials actually present in the signal. This number ($K$) is twice the number of *real* decaying sinusoids (modes). Priorto the modal analysis itself, an estimate of this number is obtained by means of the ESTER (ESTimationERror) technique[4]which consists in minimizing the error on the rotationalinvariance property of the signal subspace spanned by the sinusoids.

The block diagram in Fig.4 describes the three main steps of the high-resolution method:(a) reconstruction of the acceleration impulse response, (b) signal conditioning (band-pass filtering, downsampling…), (c)order detection, and (d) determination of modal parameters.

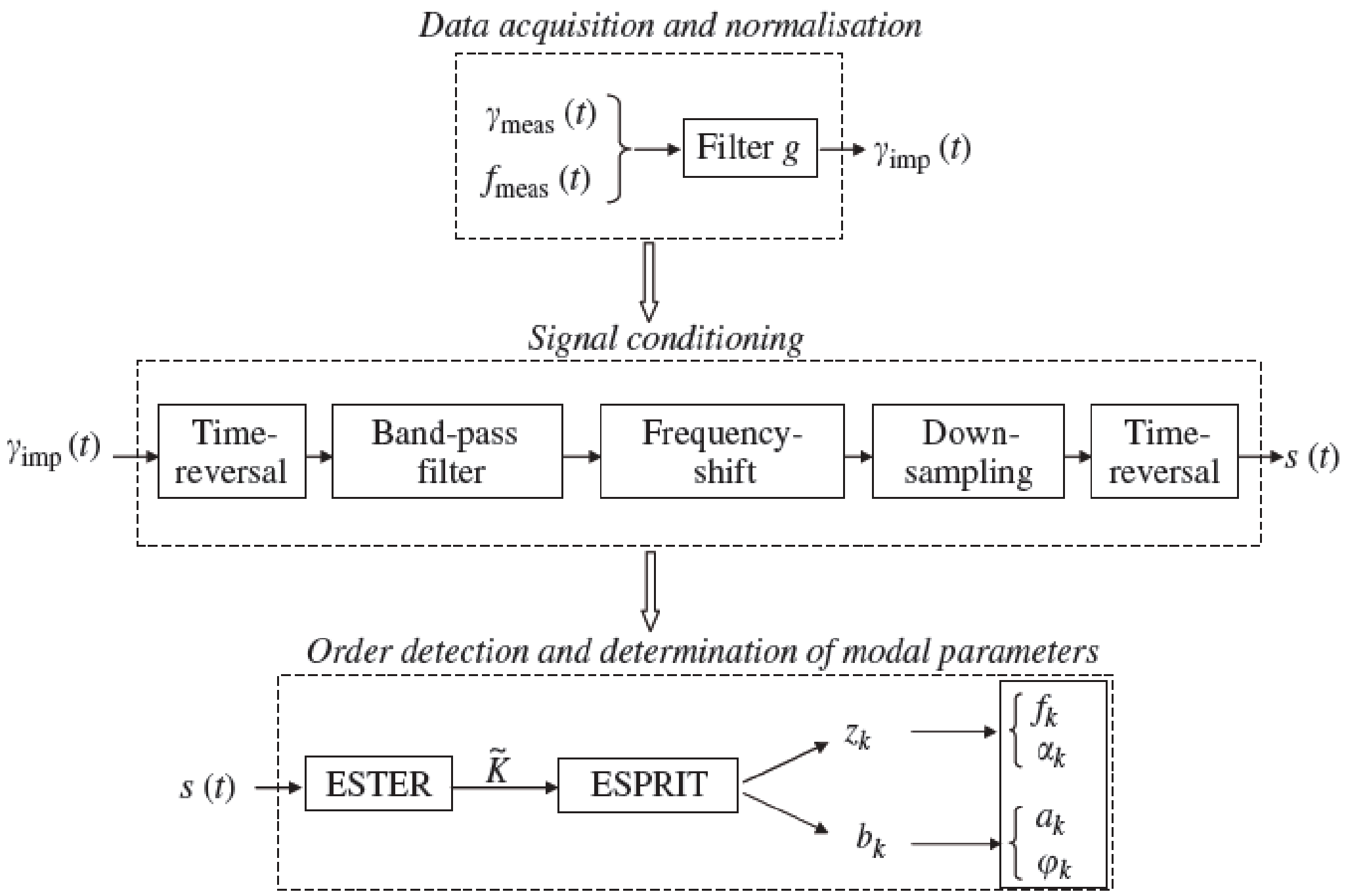

*Fig.4 - Block diagram of the high-resolution modal analysis method (see Egeet al[2] for details)*

The signal conditioning procedure -proposed by Laroche[5]- consists in splitting signals into several frequency-bands: this reduces the number of (sub-)signal components to be estimated by ESPRIT within reasonable limits and is achieved by filtering the impulse response. When narrow subbands are chosen, noise-whitening usually becomes unnecessary. The next conditioning steps aim at reducing the length of each subband signal in order to keep the memory allocation low enough and the algorithm tractable in practice: each subband signal is frequency-shifted toward zero and down-sampled.

More details on the different steps of the method are given in Ege et al[2], where the method is validated on measured and synthesized signals for frequency domain where the Fourier transform meets its limits (due to high modal overlap or poor signal-to-noise ratio).

## 3 RESULTS

For each plates under study (one aluminum plate and four bilayer plates), the modal frequencies and loss factors were estimated accurately using the method described above. A typical bank-filtering analysis is given in Fig.5 for the bilayer plate 4 (2mm of aluminum + 2mm of HDPE) where results for one narrow subband with seven highly-damped resonances are plotted.

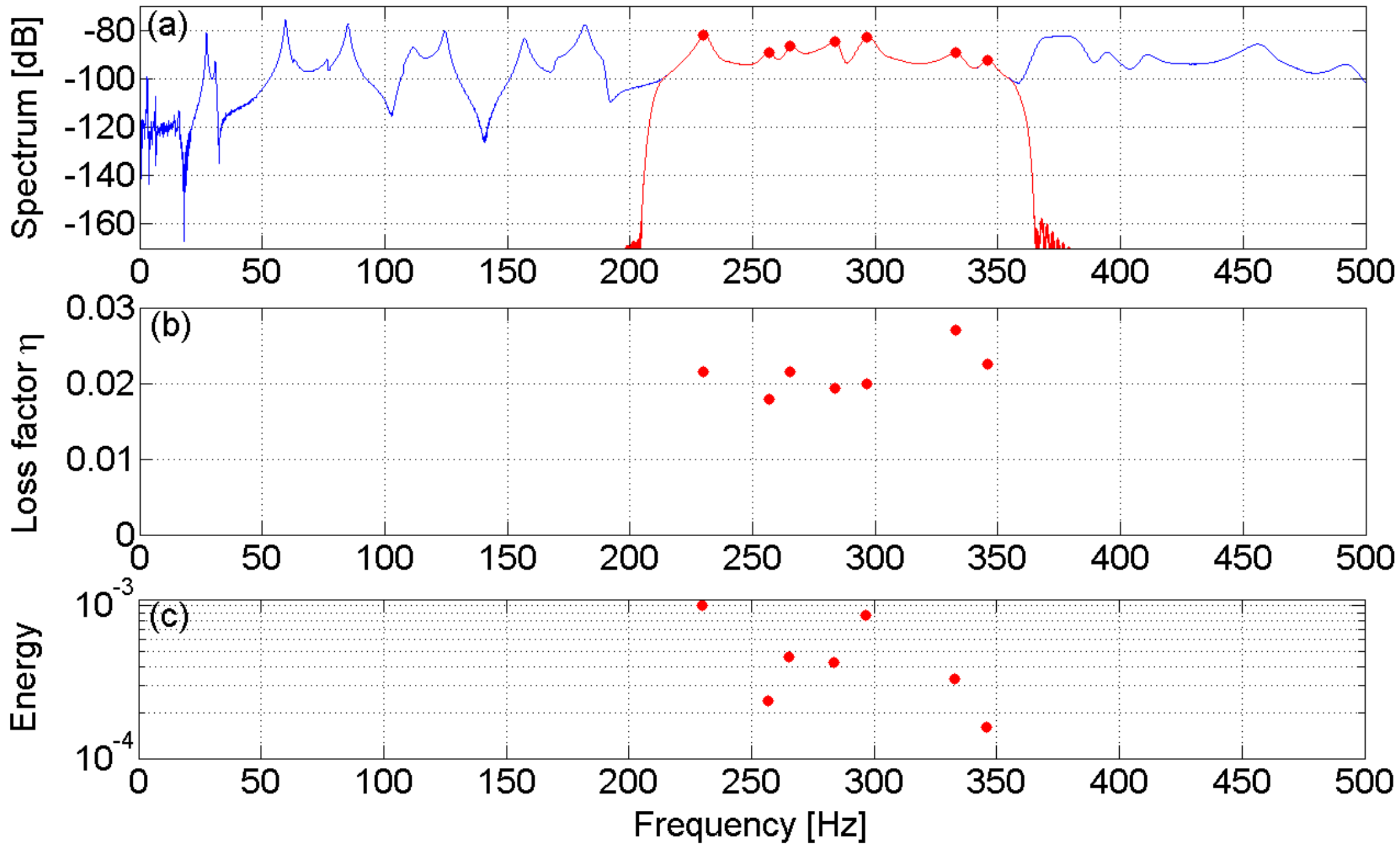

*Fig.5. Typical bank-filtering analysis of an impulse response of the bilayer plate 4 between 200 and 350 Hz. (a) —: Fourier spectrum of the impulse response; —: amplitude response of a narrow-band filter.● marks: modes estimated by the high-resolution modal analysis (modes amplitudes and frequencies). (b) ● marks : measured loss factors η versus frequencies for the seven real modes (the number of complex exponentials K=14(=2×7) is estimated with ESTER criterion). (c)● marks :Energy of each component.*

### 3.1 Loss factors

The loss factors have been estimated accurately up to 3 kHz. The Fig.6 plots the evolution with frequency of $\eta$ for the aluminum plate and the four bilayers, and Table 1 gives the mean value for each plate. These experimental estimations are compared with theoretical values obtained from two analytical models (see section 3).

### 3.2 Equivalent Young's modulus

We have estimated theequivalent Young's modulus for each plate by comparing the 13–23rd measured modal frequencies to those given by finite-element simulations. These particular modes (in the 200-400Hz subband) are chosen because they are (relatively) well-separated and the free–free boundary conditions are well-ensured. Minimizing the average of the absolute values of the relative frequency differences between experiments and FEM simulations yields $E_{ex}$ given in Table 2 (in the FEM simulations the density of the equivalent plate is fixed to the measured value for each plate, and the Poisson's ratio $\nu$ set to 0.3). With these estimated values and for each plate, the average relative frequency differences between FEM and measurements in this frequency band is less than 2%.

## 4 COMPARISON WITH MODELS

Measurements were compared with two analytical models briefly presented in this part.

### 4.1 Model A (MOVISAND)

The first model (A) calculates the equivalent complex Young's modulus $\bar{E} = E\,(1 + i\,\eta)$ of a multi-layer viscoelastic plate. Developed by Guyader*et al.*[6]and implemented in the software MOVISAND[7], the method is based on the travelling wave approach applied to a simplified multi-layer model. In each layer, bending, membrane and shear effect are considered; continuity conditions on displacement and shear stresses at interface are used to obtain the equations of motion. Hence, the method determines equivalent single layer plate material in order of having same transverse displacement.

Analytical results compares well with measurements (see Fig.6 and Table 1 for the loss factor comparison; Table 2 for the Young's modulus).The experimental $E_{ex}$ and theoretical $E_{th}$ equivalent Young's moduli match closely except for the thicker bilayer (2mm of HDPE) where the model under-estimates the value of $E$ with an error of -11%. Concerning the loss factor, the comparison may certainly be well improved if the frequency evolution of the internal loss of the HDPE was accurately measured and entered in the model. This would be done in a future work. Nevertheless, the evolution of internal loss with damping layer thickness is comparable and both estimations (experimental and theoretical) show that the viscous damping almost doubles from $\eta \approx 1.10\%$ for the plate 3 with 1.5mm of HDPE to $\eta \approx 2.13\%$ for the plate 4 of 2mm of HDPE, demonstrating the validity of the approach.

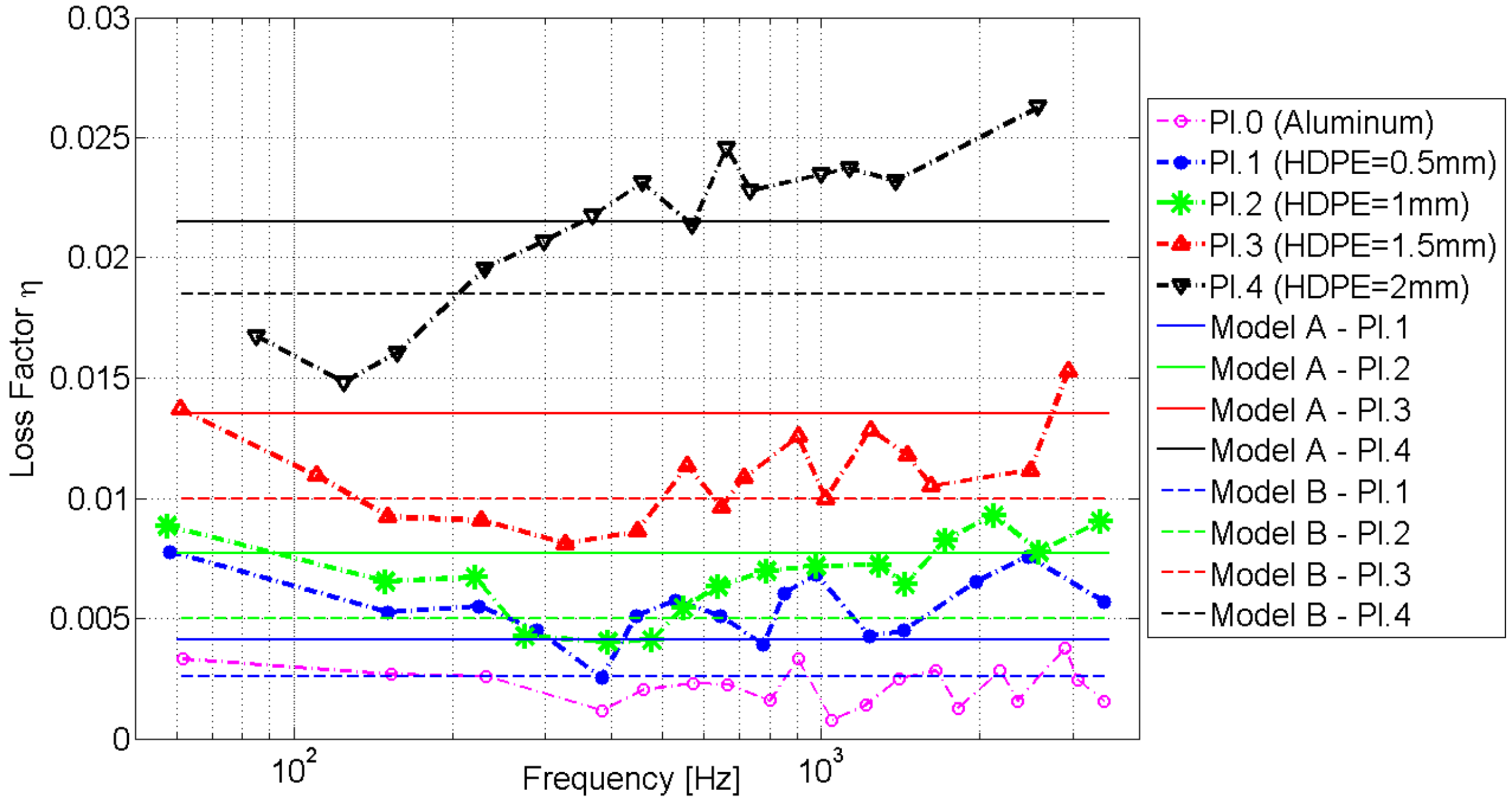

*Fig. 6– Comparison of the loss factor estimations (measurements) with predictions (analytical models) for the different bilayer plates (Pl.1 – Pl.2 – Pl.3 –Pl. 4) and the plate Pl.0 (aluminum plate of 2mm of thickness without HDPE layer). Experimental estimations of $\eta$ correspond to the mean value on each narrow subband considered.*

*Table 1 – Estimated and predicted loss factor for the five plates.*

| Plate | Measured loss factor (mean value) | Predicted loss factor Model A | Predicted loss factor Model B | Error (Model A) | Error (Model B) |
|---|---|---|---|---|---|
| Plate 0 | 0.22% | - | - | - | - |
| Plate 1 | 0.54% | 0.41% | 0.26% | -24.1% | -51.8% |
| Plate 2 | 0.68% | 0.77% | 0.50% | +13.2% | -26.5% |
| Plate 3 | 1.10% | 1.35% | 1.00% | +22.7% | -9.1% |
| Plate 4 | 2.13% | 2.15% | 1.85% | +0.9% | -13.2% |

*Table 2 – Estimated and predicted equivalent Young's modulus for the five plates.*

| Plate | Equivalent density ρ [kg.m$^{-3}$] | Estimated Young's modulus $E_{ex}$ [GPa] | Predicted Young's modulus $E_{th}$ [GPa] Model A | Error |
|---|---|---|---|---|
| Plate 0: Aluminum (2mm) | 2790 | 72 | (72) | - |
| Plate 1: Aluminum (2mm) + HDPE (0.5mm) | 2421 | 37.2 | 38 | +2.1% |
| Plate 2: Aluminum (2mm) + HDPE (1mm) | 2175 | 22.6 | 21.8 | -3.7% |
| Plate 3: Aluminum (2mm) + HDPE (1.5mm) | 1999 | 16 | 15 | -6.7% |
| Plate 4: Aluminum (2mm) + HDPE (2mm) | 1868 | 11.7 | 10.4 | -11.1% |

### 4.2 Model B (Solving Navier's equations)

Measurements were also compared with a second approach (model B) where the layer motion is described by means of the three-dimensional Navier equations. Developed by Laulagnetet al[8], the displacement solutions of Navier's equations are expressed using asymptotic expansions in the layer thickness (allowing therefore the thickness deformation). Predicted loss factor are given in Fig.6 and Table 1. Here again, the model B predicts as for model A and experiments the same evolution of internal loss with thickness but seems to under-estimates systematically the damping value.

## 5 DISCUSSION AND CONCLUSION

Experimental estimations on bilayer plates with several thicknesses of thermoplastic layer are given and compared with two models, demonstrating the validity of the approach. The equivalent Young's moduli and loss factors have been estimated through modal analysis. In the context of lightweighting, this study is meaningful as long as it shows that a 2mm HDPE layer on a 2 mm aluminum plate multiply by more than 10 the damping of the aluminum plate with only 30% of added mass.

*In perspective of this work, and in order to improve the comparison of experimental results with analytical estimations, the accurate knowledge of the frequency evolution of the complex*

*Young's modulus of the HDPE is crucial. To this end, the determination William*-Landel-Ferry (*WLF*) type plot appears to be indispensable. This would be done in a future work.

Finally, this kind of work highlights the performances of models as MOVISAND developed at LVA. With this tool, optimization of multi-layer plates with optimized damping properties for given frequency domains may be easily performed.

## 6 ACKNOWLEDGEMENTS

We acknowledge Ha Dong Hwang for the help during measurements.